\documentclass[%
 reprint,
superscriptaddress,
 amsmath,amssymb,
 aps,
 prd, 
]{revtex4-2}

\RequirePackage{graphicx}
\usepackage{amsmath}
\usepackage{amsfonts}
\usepackage{amsthm}
\usepackage{mathrsfs}
\usepackage{amssymb}
\usepackage{epsfig}
\usepackage{hyperref}
\usepackage{color}
\usepackage{amscd}
\usepackage{xcolor}

\def\prd{Phys. Rev. D }

\def\mnras{Monthly Notices of the Royal Astronomical Society }
\def\apj{The Astrophysical Journal }
\def\aap{Astronomy and Astrophysics }

\def\pasj{Publications of the Astronomical Society of Japan }
\def\apjl{Astrophysical Journal Letters }

\newcommand{\beq}{\begin{equation}}
\newcommand{\eeq}{\end{equation}}
\newcommand{\bea}{\begin{eqnarray}}
\newcommand{\eea}{\end{eqnarray}}

\def\LL{{\cal L}}

\graphicspath{{pic/}}

\def\Opava{Research Centre for Theoretical Physics and Astrophysics, Institute of Physics, Silesian University in Opava, CZ-74601 Opava, Czech Republic}
\def\Almaty{Institute of Experimental and Theoretical Physics, Al-Farabi Kazakh National University, Almaty 050040, Kazakhstan}
\def\Kazakh{Department of Physics, Kazakh National Women’s Teacher Training University, 050000 Almaty, Kazakhstan}

\begin{document}

\title{Interaction of Black Hole Magnetospheres with Inclined Ambient Fields
}

\author{Madina Zhakipova}\email{zhakipovamadina@gmail.com}
\affiliation{\Almaty}
\author{Arman Tursunov}\email{arman.tursunov@physics.slu.cz}
\affiliation{\Opava}
\author{Saken Toktarbay}\email{toktarbay.saken@kaznu.kz}
\affiliation{\Almaty}\affiliation{\Kazakh}
\author{Martin Kolo\v{s}}\email{martin.kolos@physics.slu.cz}
\affiliation{\Opava}

\begin{abstract}

Magnetic fields play a central role in black hole astrophysics, powering relativistic jets and other energetic phenomena. 
While near-horizon magnetic field is usually assumed to originate from the accretion flow, additional large-scale magnetic fields—such as those supplied by a companion neutron star in stellar-mass binaries or by galactic fields around supermassive black holes—may also affect the horizon-threading flux. 
In this work, we study the superposition of a weak arbitrarily inclined external uniform magnetic field with the internal Blandford–Znajek  split-monopole field around a Schwarzschild black hole. This setup generically gives rise to magnetic null points, where the total field vanishes. We compute the magnetic flux through an arbitrarily tilted hemisphere of the event horizon and show that the flux can be substantially suppressed by the external field. In the axisymmetric case, the flux can even vanish completely. However, with nonzero inclination, complete cancellation becomes impossible, despite significant reduction. 
We further explore the ionization and subsequent particle acceleration from a Keplerian accretion disk, finding that efficient collimated outflows persist even under significant field inclination. 
We show that the acceleration is critically dependent on the external field orientation, with the escape fraction maximized at non-zero inclinations due to the destabilization of trapping zones and  minimized in the anti-aligned configuration, where closed magnetic loops effectively suppress the outflow. 
We discuss the astrophysical implications of these findings, proposing that geometric flux cancellation can serve as a mechanism for jet quenching in compact binaries and offering an explanation for the lack of a prominent large-scale jet in Sgr A*.

\end{abstract}

\maketitle

\section{Introduction}

The structure of black hole magnetospheres remains one of the key yet unresolved problems in black hole astrophysics. Despite their critical role in launching relativistic jets, driving accretion dynamics, and shaping multimessenger signals, the exact configuration of electromagnetic fields in the vicinity of astrophysical black holes is still uncertain. This uncertainty stems from both theoretical limitations and observational constraints: while general relativistic magnetohydrodynamic (GRMHD) simulations offer insights into jet formation, they often assume idealized magnetic field geometries and boundary conditions. On the observational side, even with breakthroughs such as the Event Horizon Telescope \cite{2021ApJ...910L..13E,2024ApJ...964L..26E}, we can only indirectly infer the magnetospheric structure through polarimetric signatures and jet morphology on larger scales.

In this paper, we argue that magnetospheres are expected to arise from a combination of internal and external field sources. Internally, magnetic fields are thought to be generated by the accretion flow in strong gravitational field of the black hole. These, internal fields are central to mechanisms like the Blandford–Znajek (BZ) process \cite{Bla-Zna:1977:MNRAS:}, which taps into the rotational energy of the black hole to power relativistic jets. However, this picture neglects the possible influence of ambient magnetic fields originating from the larger astrophysical environment—such as the interstellar medium, galactic fields, or magnetized companion objects in binary systems. 

In realistic settings, these external magnetic fields are neither negligible nor necessarily aligned with the black hole spin axis. Their interaction with the internal fields can introduce significant topological complexity: magnetic flux may be suppressed through destructive interference, field lines may form closed loops or null points near the horizon, and reconnection events can occur where the fields are oppositely directed. Such features can critically affect energy extraction, jet collimation, and plasma dynamics in the innermost regions. 

In this paper, we investigate the superposition of two magnetic field configurations near a black hole: the axisymmetric split-monopole field associated with the BZ mechanism, and an inclined uniform magnetic field representing an external, ambient contribution. This work generalizes our previous study of axisymmetric field superposition \cite{2024PhRvD.109f3005K} by introducing a nonzero inclination of the external field, thereby capturing the generic case of misaligned ambient magnetization. 
Working within the Schwarzschild spacetime as a tractable test case, we analytically compute the resulting field configuration, identify the emergence of magnetic null points, and study how the magnetic flux through the horizon is affected by the relative orientation and strength of the two components and jet launching mechanisms.

The paper is organized as follows. In Section~\ref{sec:1}, we introduce the magnetic field configuration obtained by superposing an axisymmetric Blandford–Znajek split-monopole with an inclined uniform external field given by Bičák–Janiš solution in Schwarzschild spacetime. Section~\ref{sec:flux} examines the resulting magnetic flux threading the event horizon. In Section~\ref{sec:ion_Kep}, we study the ionized Keplerian accretion disks in this configuration and consequent escape and collapse of charged particles originating from the disk. In Section~\ref{sec:astro}, we discuss the astrophysical implications of our model for two representative classes of systems: stellar-mass black hole binaries (Galactic microquasars) and supermassive black holes (SMBHs), with particular emphasis on Sgr~A*. In  
Section~\ref{sec:dis_con} we summarize our main results concluding the paper. We mostly use geometrized units with  
$c=1=G$; however, for astrophysically relevant parts, the estimates are given in Gaussian units.

\section{Magnetic field combination} \label{sec:1}

\begin{figure*}
  \centering
  \includegraphics[width=0.96\textwidth]{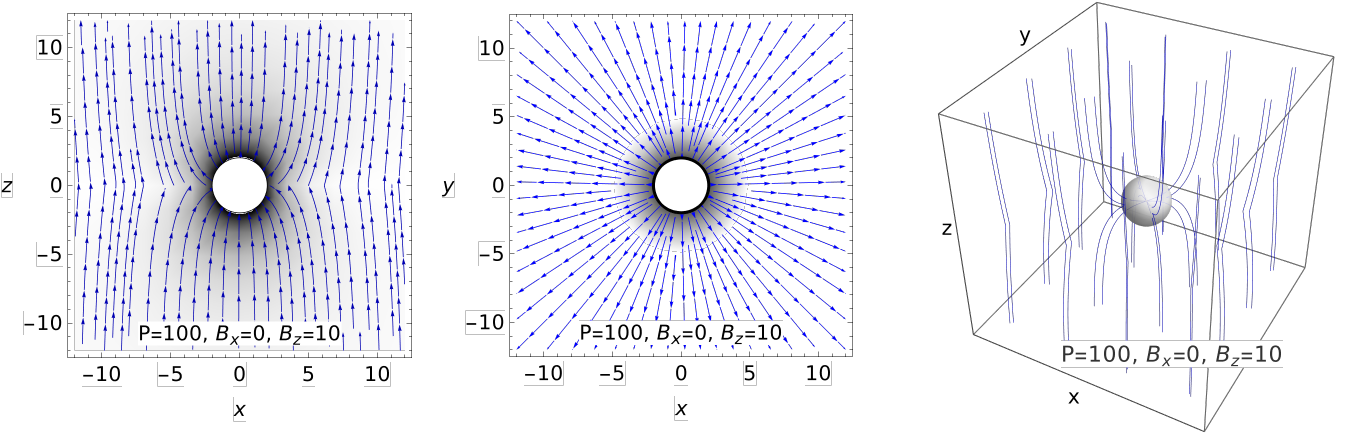}%
  \vspace{-0.1cm}
  \includegraphics[width=0.96\textwidth]{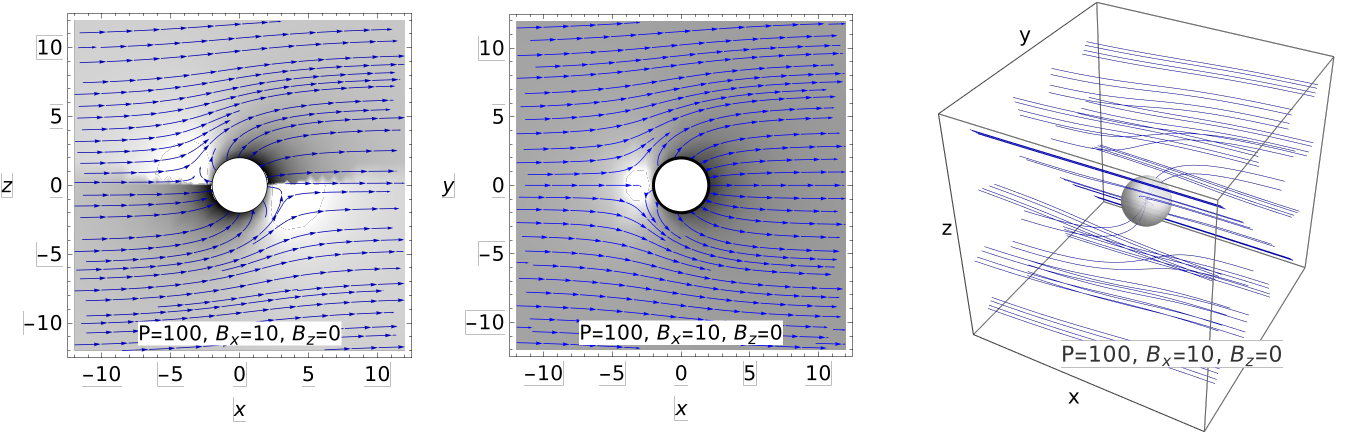}%
  \vspace{-0.1cm}
  \includegraphics[width=0.96\textwidth]{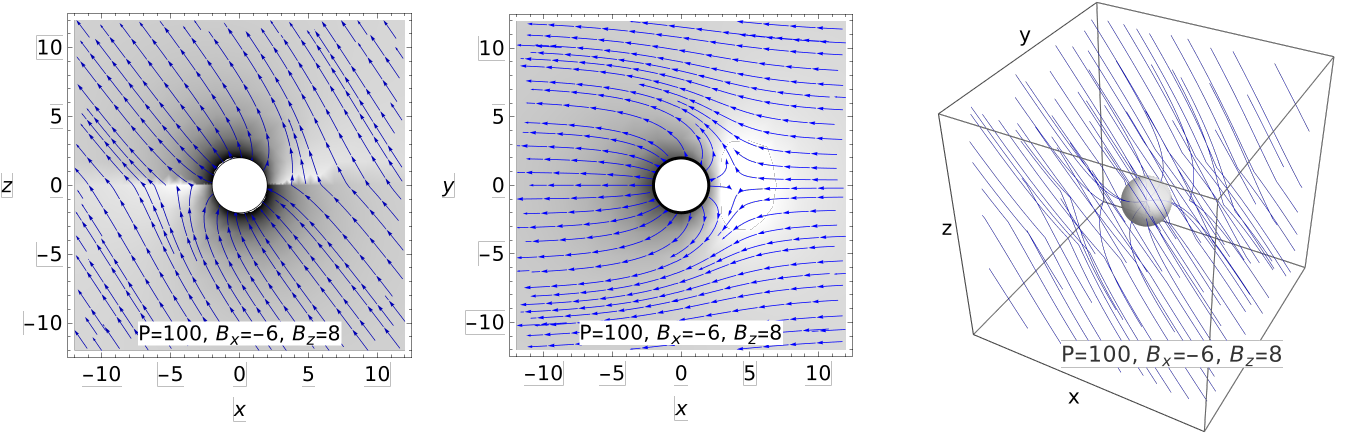}%
  \vspace{-0.1cm}
  \includegraphics[width=0.96\textwidth]{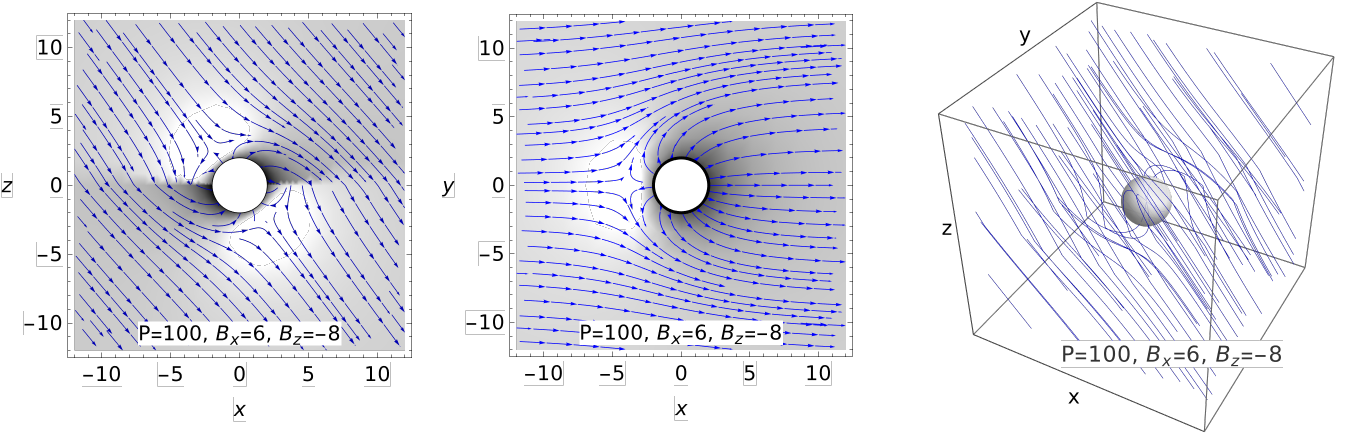}%
  \caption{ 
  Magnetic field structure for various inclination angles of the external magnetic field, while keeping the ratio $P/\sqrt{B_x^2+B_z^2} = 10$ fixed. Each row corresponds to a different inclination. The left and center panels show 2D cross-sections in the $\phi = 0$ plane, with arrows indicating field direction and shading representing magnetic field strength (field magnitude). The right panels show the corresponding 3D structures. Inclination introduces azimuthal asymmetry, field line distortion, and in some cases the formation of null points and closed loops. }
  \label{fig:1}
\end{figure*}

We start with the Schwarzschild metric describing a black hole of mass $M$ 
\beq
ds^2 = -f(r) dt^2 + f^{-1}(r) dr^2 + r^2( d\theta^2 + \sin^2\theta d\phi^2), \label{SCHmetric}
\eeq
with the lapse function given by $ f(r) = 1 - {2 M}/{r}$.

A widely used idealized configuration for modeling black hole magnetospheres is the split-monopole (SM) solution, originally proposed by \cite{Bla-Zna:1977:MNRAS:} in the context of energy extraction from rotating black holes. In this work, we consider the Schwarzschild limit of the SM configuration, where the four-vector potential has only a single non-zero component, which can be written in the form
\begin{equation}
A_{\phi}^{\rm SM} = - P |\cos{\theta}|, \label{Asmonopole}
\end{equation}
where $P$ is a constant characterizing the strength of the SM magnetic field lines threading the black hole horizon. This configuration describes a radial magnetic field emerging from the northern hemisphere and entering the southern hemisphere, with a current sheet in the equatorial plane. This ensures the antisymmetry of the field across the equatorial plane, making the solution axially symmetric, unlike the magnetic monopole solution, which is fully spherically symmetric. The SM solution carries no magnetic charge, as the magnetic field satisfies $\nabla \cdot \textbf{B} = 0$ everywhere. While idealized, this model captures essential features of jet-producing magnetospheres and is widely adopted in both analytic and numerical studies of the BZ mechanism (see, e.g. \cite{Kom:2004:MNRAS:,Tch-Nar-McK:2010:APJ:,refId0}).

In many astrophysical scenarios, the magnetic field generated by the accretion flow is not the only field present near the black hole. For example, in compact binary systems where the companion is a magnetized neutron star, the external field from the companion may become significant in the vicinity of the black hole. Similarly, SMBHs are typically embedded in large-scale galactic magnetic fields. Although these ambient fields are generally weaker than those produced by the accretion disk, they can become dynamically important on larger spatial scales, particularly in shaping the propagation and stability of relativistic jets. 

In our previous work \cite{2024PhRvD.109f3005K}, we studied the superposition of an internal SM field with an external uniform magnetic field aligned along the symmetry axis. We found that the resulting configuration may take the form of an extended paraboloidal field structure when the internal and external components are aligned, or a closed loop-like configuration when the fields are anti-aligned. In the latter case, magnetic field lines emerging from the black hole are forced to close within a finite region, rather than extending to infinity, thereby limiting the possibility of sustained jet propagation.

However, in realistic cases, external magnetic fields are unlikely to be perfectly aligned with the black hole spin or the accretion axis. These fields are typically generated by distant, uncorrelated plasma sources, and may therefore be inclined with respect to the black hole symmetry axis. To capture this generic and more realistic situation, we consider here the superposition of the internal SM field with an external magnetic field inclined at an arbitrary angle. For the external component, we adopt the exact solution by \cite{1985MNRAS.212..899B}, which describes a uniform magnetic field at an arbitrary inclination around black hole. 

The total four-potential thus represents a superposition of an axisymmetric BZ split-monopole field with an inclined external field described by the Bičák–Janiš solution. The combined nonzero components read   
\begin{align}
A_{r}  &= - \,\frac{B_x}{2} (r-M) \sin2\theta \, \sin\phi,\\
A_{\theta}&=-B_x (r^2 \cos^2\theta - M r \cos2\theta ) \sin\phi,\\
A_{\phi} &= 
- P |\cos{\theta}| + \frac{B_z}{2} r^2 \sin^2\theta \nonumber \\ 
&- \frac{B_x}{2} (r-M) r \sin2\theta \cos\phi .
 \label{Acombined}
\end{align}
Here, $B_z$ and $B_x$ denote the vertical and horizontal components of the external uniform magnetic field, respectively, while $P$ characterizes the strength of the internal SM configuration. The angle between the external field and the black hole’s symmetry axis is characterized by the inclination angle $i$, defined by  
$\tan i = B_x/B_z$.  

The radial and polar components $A_r$ and $A_\theta$ arise solely from the inclined external field and vanish in the axisymmetric limit $i = 0$. In this case, the solution reduces to the aligned configuration analyzed in our previous work \cite{2024PhRvD.109f3005K}. For nonzero inclination, the magnetic field structure becomes significantly more complex, as we demonstrate below.

\subsection{The field components}

The orthonormal components of the magnetic field vector  
$\mathbf{B} = \left(B_{\hat{r}}, B_{\hat{\theta}}, B_{\hat{\phi}}\right)$,  
as measured by a locally non-rotating observer (often referred to as a zero-angular-momentum observer, or ZAMO), are given by  
\begin{equation}
B_{\hat{i}} = \eta_{ijk} \sqrt{g^{jj} g^{kk}}\, F_{jk},
\end{equation}  
where \( \eta_{ijk} \) is the Levi-Civita symbol in three-dimensional space, \( g^{jj} \) are the components of the inverse spatial metric in Schwarzschild coordinates, and \( F_{jk} \) is the electromagnetic field tensor.  
In Schwarzschild spacetime, orthonormal basis vectors of the ZAMO frame are given by  
\begin{equation} 
\begin{split}
(e_{\hat{0}}, e_{\hat{1}}, & e_{\hat{2}}, e_{\hat{3}}) = \\ 
& \left( f^{-1/2} \partial_t,\ f^{1/2} \partial_r,\ r^{-1} \partial_\theta,\ (r\sin\theta)^{-1} \partial_\phi \right),
\end{split}
\end{equation} 
where $f(r) = 1 - {2M}/{r}$. 
The non-vanishing orthonormal components of the magnetic field measured by the ZAMO take the form:
\begin{align}
B_{\hat{r}} &= \frac{P\, |{\cos\theta}| \sec\theta}{r^2} + B_z \cos\theta + B_x \cos\phi \sin\theta ,  \label{Brhat} \\
B_{\hat{\theta}} &= \sqrt{f(r)} \left( B_x \cos\theta \cos\phi - B_z \sin\theta \right), \label{Bthetahat} \\
B_{\hat{\phi}} &= -B_x \sqrt{f(r)} \sin\phi. \label{Bphihat}
\end{align}
In terms of coordinate basis components, the magnetic field vector is expressed as: 
\begin{align} 
B_r &= \frac{\sqrt{f(r)}}{r^2} \big( P\, |{\cos\theta}| \sec\theta  + r^2 B_z \cos\theta \nonumber \\
& + r^2 B_x \cos\phi \sin\theta \big),  \\
B_\theta &= \frac{\sqrt{f(r)}}{r} \left( B_x \cos\theta \cos\phi - B_z \sin\theta \right), \\
B_\phi &= -\frac{B_x \sqrt{f(r)} \sin\phi}{r \sin\theta}.
\end{align}
These expressions describe the spatial structure of the combined magnetic field as observed in the orthonormal and coordinate frames. In particular, the  $B_{\hat{r}}$ component receives contributions from the vertical $B_z$, horizontal $B_x$, and monopolar $P$ terms, whereas $B_{\hat{\theta}}$ and $B_{\hat{\phi}}$ reflect the azimuthal asymmetry induced by the inclined external field. 

\subsection{Morphologies of the combined field}

In Figure~\ref{fig:1}, we demonstrate examples of the magnetic field lines in the combined configuration, with the strengths of the field components 
($P, B_x, B_z$) explicitly labeled in each panel. All configurations shown correspond to the same magnitude of internal field $P=100$, while varying the orientation and relative contributions of the external field components $B_x$ and $B_z$, thus altering only the inclination angle $i$. Each row corresponds to a different inclination scenario, ranging from purely vertical external field ($B_x=0$, top row) to purely horizontal ($B_z=0$, second row), and to mixed aligned or anti-aligned configurations in the lower rows. The columns show orthogonal cross-sections in the $xz$ and $yz$ planes (left and center), and a 3D visualization of the field lines (right). The background shading represents the magnetic field strength (i.e., field magnitude), and blue arrows trace the direction of the field lines. All 2D plots correspond to the azimuthal slice $\phi=0$, where the inclination-induced asymmetries are most evident. One can see that varying the inclination angle leads to qualitatively different magnetic field topologies. Below we discuss some of them:   

\textbf{(1)} 
In the perfectly aligned case (e.g. $B_x=0, B_z>0, P>0$, first row in Figure~\ref{fig:1}), the field configuration remains axisymmetric and exhibits the typical collimated paraboloidal structure, with field lines extending smoothly from the black hole to infinity. This configuration is strongly supported by GRMHD and GRPIC simulations, which consistently show the emergence of a paraboloidal magnetosphere around the black hole, largely independent of the initial magnetic field setup or accretion flow structure (see, e.g., \cite{Kom-McK:2007:MNRAS:,Tch-Nar-McK:2010:APJ:,Kol-Jan:2020:RAG:,Par-Phi-Cer:2019:PRL:}). Notably, this geometry closely resembles the large-scale jet morphology observed in astrophysical black hole systems \cite{Nak-etal:2018:APJ:,Kol-Sha-Tur:2023:EPJC:}. 

\textbf{(2)} 
In the anti-aligned configuration, where the uniform external magnetic field opposes the Blandford–Znajek split-monopole ($B_z$ and $P$ have different signs), the resulting topology exhibits closed magnetic field loops connecting the black hole to a thin equatorial disk. In the case of a rotating black hole, such magnetic linkages have long been known to enable the transfer of energy and angular momentum from the black hole to the disk through magnetic connections \citep{2000ApJ...533L.115L,2007MNRAS.374..647W}. This process can be interpreted as a disk-mediated variant of electromagnetic energy extraction, complementary to the standard energy extraction mechanisms operating along open field lines \citep{Bla-Zna:1977:MNRAS:,Tur-Dad:2019:Universe:}, and may play an important role in regulating jet power, disk heating, and variability.

\textbf{(3)} Purely horizontal external field ($B_z=0, B_x>0, P>0$, second row in Figure~\ref{fig:1}), in contrast to the above cases, breaks the axial symmetry, causing the field lines to bend in the azimuthal direction and introducing significant distortion near the black hole. In this case, the external magnetic field is oriented perpendicular to the expected jet propagation direction, potentially disrupting or dissipating the outflow due to strong magnetic deflection and the absence of collimating field lines. As we demonstrate numerically below, this configuration significantly alters the trajectories of escaping charged particles from an ionized Keplerian disk, preventing sustained jet formation. 

\textbf{(4)} In the inclined cases with aligned $B_z$ and $P$ and arbitrary $B_x\neq0$ (e.g. third row in Figure~\ref{fig:1}), the resulting field structure exhibits features of both the aligned and anti-aligned cases. Near the black hole, the field retains a paraboloidal shape, but it is noticeably asymmetric due to the horizontal component $B_x$. At the same time, the inclination breaks axial symmetry and introduces magnetic null points, typically associated with the closed loop-like configurations seen in the anti-aligned setups. This hybrid behavior shows that even moderate inclinations can qualitatively alter the near-horizon magnetic topology. 

\textbf{(5)}
In the inclined anti-aligned configuration with the opposite signs of $B_z$ and $P$, while arbitrary $B_x\neq0$ (e.g. fourth row in Figure~\ref{fig:1}), the magnetic field structure shows characteristics of the loop-like topology found in the purely anti-aligned case. Here, all field lines connecting to the black hole are redirected toward the equatorial plane, where the accretion disk is expected to reside. The presence of oppositely directed vertical components ($B_z<0$) causes partial cancellation of the fields and leads to the formation of magnetic null points above the equatorial region. At the same time, the finite horizontal component $B_x$ introduces a tilt that breaks axial symmetry and bends the field lines in the azimuthal direction. The resulting configuration is therefore hybrid: some field lines close near the black hole, while others extend outward at an angle. Such a topology is particularly favorable for magnetic reconnection and may strongly suppress or destabilize jet outflows.

Magnetic null points in the inclined configuration can be identified from the orthonormal field components (\ref{Brhat})–(\ref{Bphihat}). Their location is determined by the conditions 
\begin{align}
&r^2_{\rm null} =-\frac{P B_z|\cos\theta|}{(B_x^2+B_z^2)\cos^2\theta}, \\ 
&\tan\theta_{\rm null}  = \frac{B_x}{B_z}, \quad \phi_{\rm null}  = 0, 
\end{align}
where the last relation is valid for $B_z P < 0$. These expressions define the angular position and radial distance of the null points relative to the black hole.

Overall, these examples illustrate the strong dependence of the near-horizon field geometry on the inclination of the external field and its orientation relative to the internal axis. As we show in the following sections, such geometric distortions significantly affect the magnetic flux threading the horizon and the resulting conditions for jet formation.

\section{Magnetic fluxes across the horizon}
\label{sec:flux}

In stationary black hole magnetospheres, including both vacuum and force-free cases, the amount of magnetic flux crossing the event horizon is directly linked to how effectively rotational or orbital energy can be extracted. The presence of this flux is essential for enabling energy extraction mechanisms, such as the BZ mechanism, and may play a key role in powering relativistic jets. 

For Kerr black holes, Bičák \& Janiš \cite{1985MNRAS.212..899B} investigated how external magnetic fields behave near rotating black holes and showed that, for tilted fields,  magnetic field lines can thread the event horizon without being expelled, even in the extremal Kerr limit. They gave a geometric and gauge-invariant calculation of the flux through an arbitrarily oriented horizon hemisphere in a uniform and arbitrary inclined magnetic field. 

In this section, we adopt their original definition of the flux, applying it to our combined magnetic field model in Schwarzschild case. Instead of the Kerr geometry, an axisymmetric split-monopole magnetic field is imposed, which preserves the symmetry of the problem and allows for a consistent and directly comparable treatment.

Following \cite{1985MNRAS.212..899B}, the magnetic flux through a horizon hemisphere $\mathcal H(\alpha,\beta)$ is given by 
\begin{equation}
\begin{split}
\Phi(\alpha,\beta)
&=\int_{\mathcal H(\alpha,\beta)}\! F_{\theta\phi}\,d\theta\,d\phi  \\
&=\int_{0}^{2\pi}\! d\phi \int_{0}^{\theta_0(\phi;\alpha,\beta)}
F_{\theta\phi}(r_H,\theta,\phi)\,d\theta, \label{eq:Flux-def}
\end{split}
\end{equation} 
where $r_H=2M$ for Schwarzschild, and the hemisphere is specified by the direction of its outward normal through two angles $(\alpha,\beta)$: a tilt $\alpha$ away from the $+{\boldsymbol{z}}$ axis and an azimuth $\beta$ measured from $+{\boldsymbol{x}}$. The boundary function is given by 
\begin{equation}
\theta_0(\phi;\alpha,\beta)
=\frac{\pi}{2}+\arctan\!\Big[\tan\alpha\,\cos(\phi-\beta)\Big].
\label{eq:theta0}
\end{equation}
On the horizon $r=r_H$, the $F_{\theta\phi}$–component of the field 
is given by 
\begin{equation}
\begin{split}
F_{\theta\phi}\big|_{r_H} &=
B_x r_H^{\,2}\cos\phi\,\sin^2\theta
+ B_z r_H^{\,2}\sin\theta\cos\theta \\ 
&+ P\,|\cos\theta| \tan\theta\, .
\label{eq:Fthph-hor}
\end{split}
\end{equation}
Evaluating the $\theta$–integral in \eqref{eq:Flux-def} first, yields the following expression for the flux through an arbitrary hemisphere~$\mathcal H(\alpha,\beta)$ 
\begin{equation} 
\begin{split}
\Phi(\alpha,\beta) &=\pi r_H^{\,2}\Big(B_z\cos\alpha+B_x\sin\alpha\cos\beta\Big) \\ 
&+ 2 \, P \,(\pi-2\alpha) \,.
\label{eq:Flux-general}
\end{split}
\end{equation} 
The first two terms coincide with the Schwarzschild limit of the corresponding expression obtained in \cite{1985MNRAS.212..899B}, while the third term represents the new contribution to the flux from the split-monopole component.  

It is useful to discuss some special cases: 

\textbf{(1)} If $\alpha=0$, the hemisphere is located around the $z$-axis, yielding 
\begin{equation} \label{Eq:flux-alligned}
\Phi(0,\beta)=\pi r_H^{\,2}B_z+2\pi P, 
\end{equation} 
which is independent from $\beta$, and $B_x$, as expected. This leads to the result of purely axisymmetric flux without inclination component. One can see, that depending on the relative orientation of the external and internal field components, the flux can be set to completely zero. This occurs when the the fields have relation $P=-\frac12 r_H^2B_z$. It is important to note that this zero-flux condition does not imply a magnetic null-point at the pole (which would require $P = -r_H^2 B_z$); rather, it implies that the positive flux through the polar region is exactly canceled by the negative flux near the equator (or vice versa), creating a 'neutral ring' (or a radial null) at $\theta = 60^{\circ}$. 

\textbf{(2)} The case with $\alpha = {\pi}/{2}$, corresponds to a hemisphere centered on the equatorial plane. In this case, the expression reduces to
\begin{equation}
\Phi\left(\tfrac{\pi}{2},\beta\right) = \pi r_H^{\,2} B_x \cos\beta \, ,
\end{equation}
which depends explicitly on the azimuthal angle 
$\beta$ and the inclined external field component $B_x$. This dependence introduces a non-axisymmetric modulation in the total magnetic flux. 
If the internal component would be a pure monopole, it would contribute a constant negative offset, and a complete cancellation of the flux could occur. However, since in split-monopole the equatorial plane divides the region with ingoing and outgoing fields, the flux at the equatorial plane due to split-monopole component cancels out, leaving only the effect of the inclined field.

\textbf{(3)} For general inclination angles $\alpha$, the total flux shows a mixed dependence on both polar and azimuthal angles. The result is a superposition of an axisymmetric flux (determined by the vertical and split-monopole fields) and a non-axisymmetric component (determined by the horizontal field). 
As $\alpha$ increases from $0$ to $\pi/2$, the contribution of the axisymmetric terms decreases, while the influence of the non-axisymmetric term $B_x \cos\beta$ increases. Physically, this implies that for any oblique inclination ($\alpha \neq 0$), the magnetic flux is spatially modulated along the azimuthal direction: the value of the flux depends explicitly on the specific azimuthal orientation $\beta$ of the selected hemisphere, reaching its extrema when the hemisphere is aligned with or against the horizontal field lines.

Recent 3D general-relativistic particle-in-cell (GR-PIC) simulations \cite{refId0} provide independent numerical support for the geometric model presented here. Although their work addresses the full relativistic dynamics in a Kerr metric, their findings regarding the field topology are in good qualitative agreement with our analytical results. First, they report a significant reduction in the outgoing Poynting flux (jet power) as the magnetic field inclination increases, which mirrors our derivation showing that the net magnetic flux available to the horizon decreases for tilted configurations. Second, their simulations show that the jet structure loses its simple helical symmetry and develops complex, warped current sheets at high inclinations. This dynamic behavior corresponds directly to the strong non-axisymmetric modulation ($\propto \cos\beta$) identified in our flux formula for $\alpha \neq 0$. Finally, the 'magnetic discontinuities' they observe in the magnetosphere can be interpreted as the physical counterparts to the neutral rings (or flux cancellation zones) predicted by our superposition of split-monopole and uniform fields.

\section{Ionized particle acceleration}
\label{sec:ion_Kep}
 
 \begin{figure*}
  \centering
  \includegraphics[width=1\textwidth]{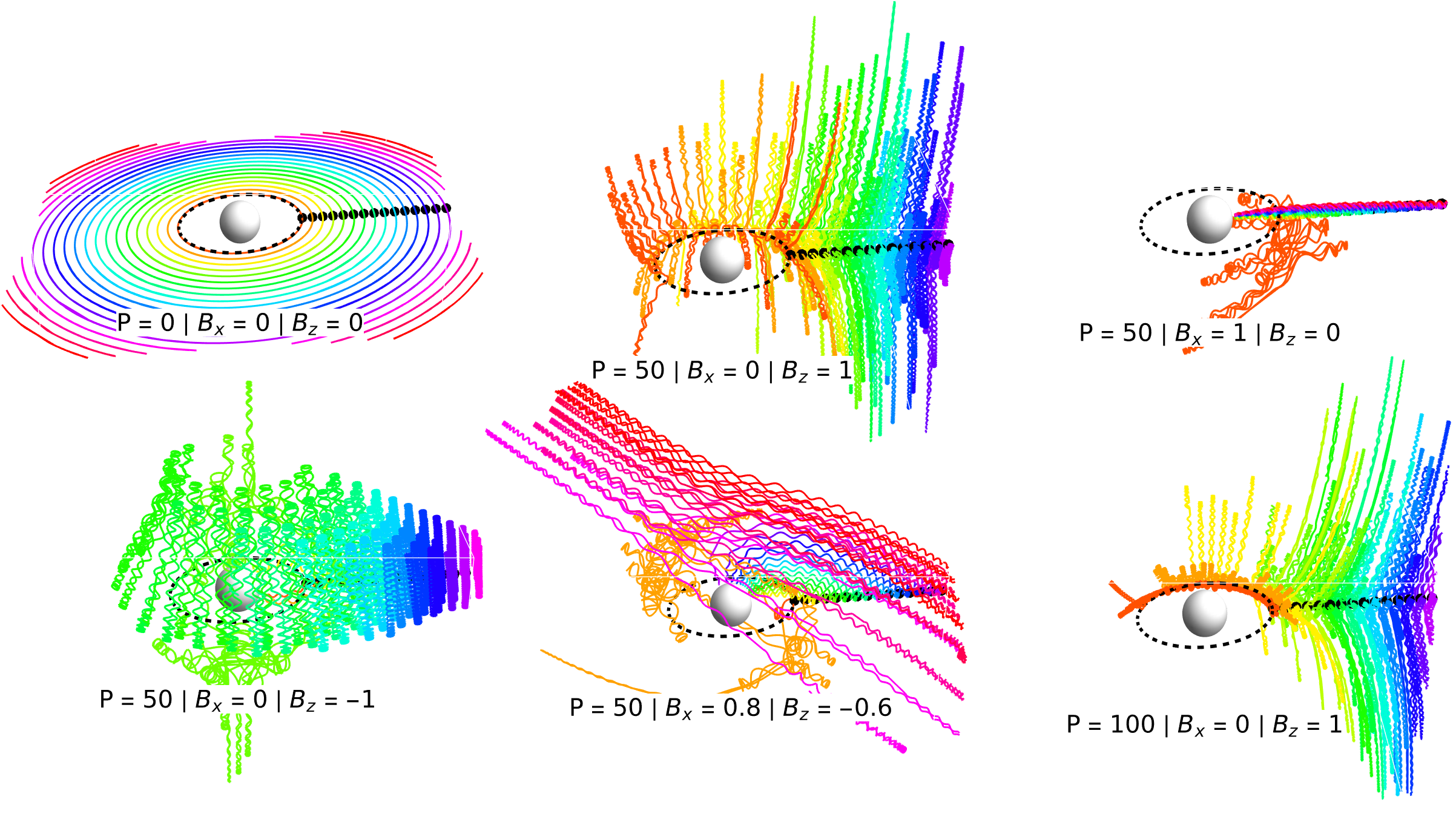}
  \caption{ Trajectories of ionized particles from thin Keplerian accretion disk initially orbiting in neutral circular geodesics with $\LL>0$ near the equatorial plane. Different colors represent different ionization position $r_0$ (indicated by black dot), but the same $\phi=0$ to better visualize the resulting motion. 
  }
  \label{fig:2}
\end{figure*}

 \begin{figure*}
  \centering
  \includegraphics[width=1\textwidth]{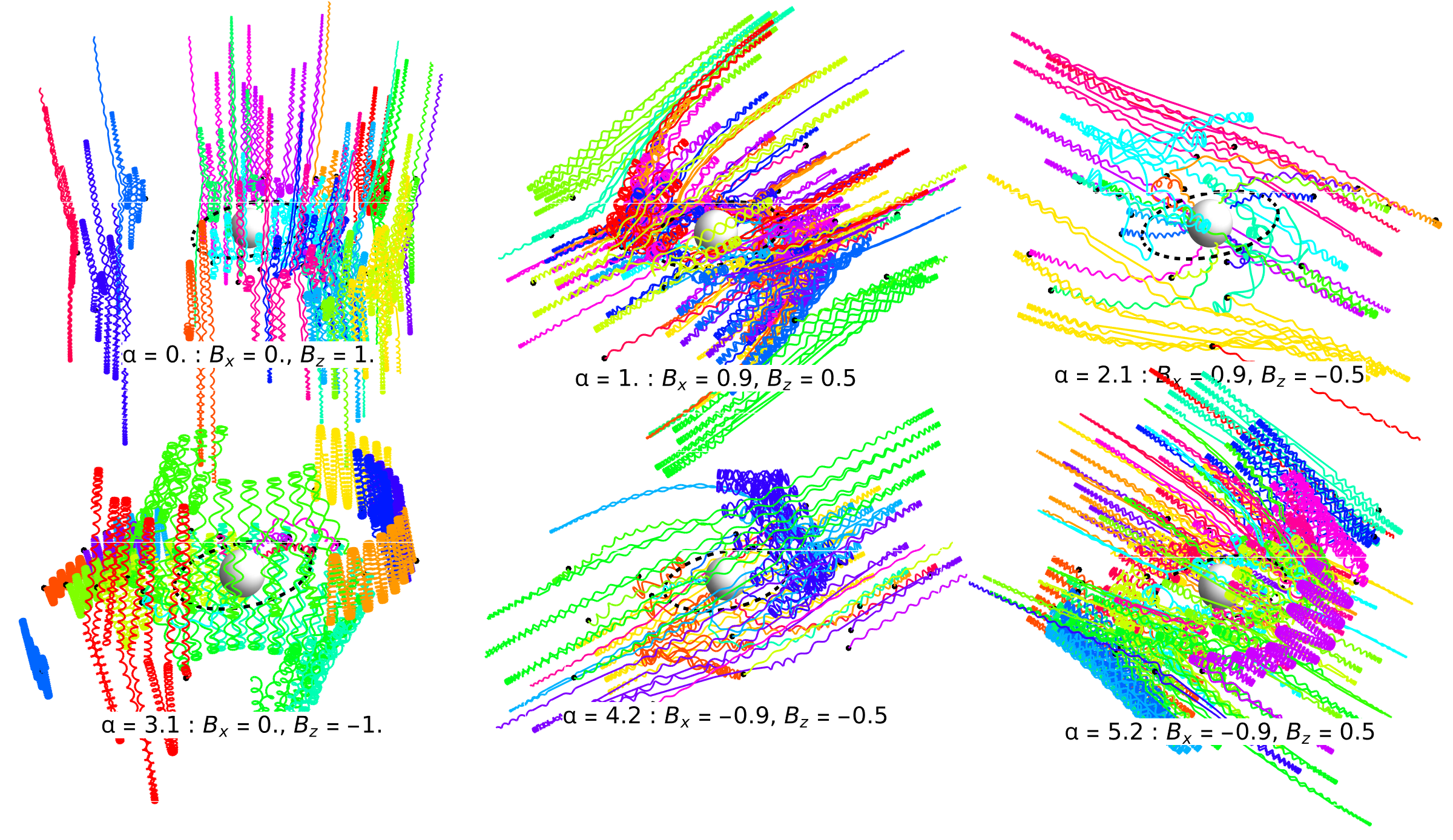}
  \caption{ 
  Same as Fig.~\ref{fig:2}, but for randomized initial phases $\phi_0 \in [0, 2\pi)$ across different $r_0$ values. Black dots indicate the ionization points. The internal field is fixed at $P=50$, and the external field inclination $\alpha$ is defined by $\tan \alpha = B_x/B_z$, with the field magnitude fixed at $|B| = \sqrt{B_x^2 + B_z^2} = 1$.
  }
  \label{fig:3}
\end{figure*}


The complex magnetic topology resulting from the superposition of the split-monopole and inclined uniform fields creates unique gradients and magnetic null regions that are critical for particle acceleration. To understand the efficiency of acceleration mechanisms in this environment, we focus in this section on the ionization of particles originating from a Keplerian accretion disk. Since we consider a non-rotating black hole, where no electric field is induced by frame dragging, the motion of ionized particles starting from the accretion disk offers an effective setting to explore magnetic field effects on the charged matter, as a useful zero-order approximation to more complex astrophysical scenarios. 

 \begin{figure*}
  \centering
  \includegraphics[width=0.8\textwidth]{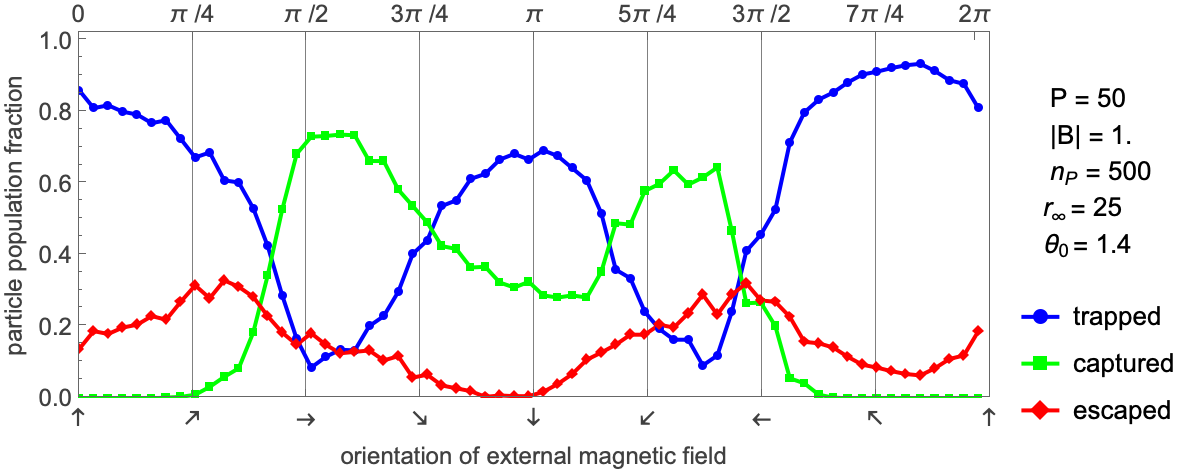}
  \caption{ Global particle population fractions (escaping, captured, and trapped) as a function of the external magnetic field inclination $\alpha \in [0, 2\pi)$. The fractions are calculated by simulating an ensemble of $n_P=500$ particles for each inclination step. Initial ionization positions are randomized uniformly across the near-equatorial ($\theta_0=1.4$ rad) disk surface, with $r_0 \in [r_{\rm ISCO}, 20]$ and $\phi_0 \in [0, 2\pi)$. 
  The escape boundary is set to $r_\infty = 25$ and magnetic field parameters: $P=50$ and $|B|=1$. }
  \label{fig:4}
\end{figure*}

In realistic astrophysical environments, accretion disks around black holes are expected to be fully ionized due to high temperatures and densities, satisfying global charge neutrality. However, local charge separation can occur at disk boundaries due to statistical fluctuations or decay processes. For example, neutrons produced in nucleosynthesis may decay into protons and electrons, initiating ionized motion from initially neutral geodesics \cite{2008ApJ...681...96H,2017Galax...5...15J,Tur-etal:2019:ApJ:}. Once separated, such particles are subject to Lorentz forces, which can alter their trajectories significantly, leading to either bounded motion, accretion, or escape.

This process was previously analyzed for axisymmetric magnetic field configurations \cite{2024PhRvD.109f3005K,Stu-Vrb:2025:Ent:}. In that case, it was demonstrated that the topology of the combined field plays a key role in determining the fate of ionized particles. Specifically: 
\begin{itemize}
    \item In the resulting paraboloidal \emph{jet-like configuration}, where both field components are aligned, open magnetic field lines provide more favorable escape channels for ionized particles than either magnetic component separately.   
    \item In the resulting \emph{loop-like configuration}, with anti-aligned components, magnetic field lines close onto the black hole, suppressing escape and favoring accretion.
\end{itemize}

The motion of the ionized particles was modeled using a simple prescription: mechanical energy and momentum are conserved during ionization, while the canonical angular momentum receives a shift due to the electromagnetic four-potential $A_\phi$. This framework allowed us to identify regions of the disk that favor escape or trapping, depending on the magnetic configuration and particle charge.

Increasing the inclination of an external magnetic field 
leads to a more complex post-ionization dynamics, as we show below. 
A neutral particle from the Keplerian disk initially follows a circular geodesics at radius $r_0$ and inclination angle $\theta_0$ with respect to the axis. Its energy and angular momentum are given by 
\begin{equation}
\mathcal{E}_{0} = \frac{r_0 - 2}{\sqrt{r_0^2 - 3r_0}}, \qquad 
\mathcal{L}_{0} = \frac{r_0 \sin\theta_0}{\sqrt{r_0 - 3}}.
\label{eq:EL_inclined}
\end{equation}
The corresponding initial conditions for the particle's position and four-momentum are 
\begin{align}
x^\alpha &= (t, r, \theta, \phi) = (0, r_0, \theta_0, \phi_0),
\label{eq:x0_inclined} \\
u_\alpha &= (u_t, u_r, u_\theta, u_\phi) = (-\mathcal{E}_0, 0, 0, \mathcal{L}_0).
\label{eq:u0_inclined}
\end{align}
After ionization, the particle begins to interact with the external electromagnetic field. Since the field is inclined, the canonical four-momentum gains contributions from all three spatial components of the four-potential. While the mechanical energy remains conserved due to the absence of a scalar potential ($A_t = 0$), the spatial components are modified as:
\begin{align}
\mathcal{E}_{(i)} &= \mathcal{E}_{(0)},
\label{eq:Epost_inclined} \\
u_r^{(i)} &= A_r(r_0, \theta_0, \phi_0),
\end{align}
\begin{align}
u_\theta^{(i)} &= A_\theta(r_0, \theta_0, \phi_0), \\
u_\phi^{(i)} &= \mathcal{L}_{(0)} + A_\phi(r_0, \theta_0, \phi_0).
\label{eq:Lpost_inclined}
\end{align}
These modifications imply that the post-ionization motion depends not only on the field strength but also on the local geometry of the magnetic field, encoded in $A_\mu(r_0, \theta_0, \phi_0)$.

In Figures~\ref{fig:2} and \ref{fig:3}, we present the trajectories of charged particles ionized from circular Keplerian orbits in the presence of a combined inclined magnetic field for various values of $B_x, B_z$, and SM parameter $P$. 
In contrast to the axially-symmetric case \cite{2024PhRvD.109f3005K}, it is clear that the particle's fate is sensitive to its precise location on the disk at the moment of ionization. In Fig.~\ref{fig:2}, we plot representative trajectories for a fixed ionization phase $\phi_0 = 0$. 
In Fig.~\ref{fig:3} we randomize the initial ionization phase $\phi_0$ uniformly over the interval $[0, 2\pi)$ for each radius $r_0$.  These figures show that the nonzero component $B_x$ breaking the axial symmetry, induces complex vertical and radial oscillations that are absent in the aligned field configuration.

The fate of each particle can be classified into one of three distinct populations: \textit{(i) escaping particles}, which reach a large distance from the black hole; \textit{(ii) captured particles}, which cross the event horizon; and \textit{(iii) trapped particles}, which remain on bounded orbits within the simulation domain. To quantify how the global acceleration efficiency depends on the geometry of the external field, we simulate a large ensemble of test particles ionized from the equatorial Keplerian disk at various initial radii $r_0$ and randomized $\phi_0$. 

Figure~\ref{fig:4} illustrates how the relative distribution of these populations changes as we rotate the external magnetic field vector with respect to internal field $P$. We performed a high-resolution parameter sweep over the magnetic inclination angle $\alpha$ (where $\tan \alpha = B_x/B_z$), covering the full range $\alpha \in [0, 2\pi)$. For each value of $\alpha$, we simulated an ensemble of $n_P=500$ particles with initial positions randomized uniformly across the disk surface: radial coordinates were taken from the range $r_0 \in [r_{\rm ISCO}, 20]$ and azimuthal angles from $\phi_0 \in [0, 2\pi)$. 

The simulation procedure for each particle follows the following scheme. First, a neutral seed particle starts on a near-equatorial circular Keplerian geodesic at $r_0$. It is allowed to evolve until it reaches the randomly selected ionization phase $\phi_0$, at which point the ionization event occurs. The canonical momentum is then updated according to the local vector potential as described above in this section. The charged trajectory is then integrated numerically up to a maximum proper time of $\tau_{\max} \approx 800$. If the particle reaches the outer boundary before this time (in our case we set $r_\infty=25$), it is classified as escaping; if it falls below the horizon radius, it is captured; otherwise, it is counted as trapped. The resulting plot reveals the global effectiveness of the acceleration mechanism as a function of the field geometry and inclination angle.

Below we list several key features of particle dynamics in the combined inclined magnetic field configuration based on our numerical tests: 
\begin{itemize}

\item The global acceleration efficiency is strongly modulated by the inclination angle $\alpha$ of the external field. Our Fig.~\ref{fig:4} reveals that specific field orientations maximize the population of escaping particles, while others suppress it. This implies that in addition to intrinsic black hole and disk parameters, the large-scale external magnetic field geometry can play a crucial role in regulating the particle outflow. 

\item Interestingly, the axisymmetric jet-like configuration ($P>0$, $B_z>0$, $B_x=0$), while topologically optimal for jet collimation, is not the most efficient for particle acceleration. We observe that the escape fraction is actually maximized by a slight inclination of the external field rather than perfect alignment. 

    \item 
       Unlike axisymmetric configurations, the particle dynamics become highly sensitive to the initial azimuthal position $\phi_0$. We observe that particles ionized at the same radius but different phases can follow diverging trajectories and fates. 
    
    \item The transition from "trapped" to "escaping" regime is mediated by a chaotic scattering. The inclined field breaks the conservation of vertical angular momentum, allowing particles to exchange energy between vertical and radial oscillations. 

    \item Certain configurations with oppositely directed vertical and horizontal magnetic field components, nominally expected to resemble the loop-like case, can instead lead to collimated outflows. This indicates that the loop-like regime is no longer topologically stable in the inclined case and may transition into a jet-like configuration depending on the relative strength and orientation of the field components. 

    \item In a purely vertical anti-aligned external magnetic field ($B_z < 0, B_x=0$), the configuration reduces to the axisymmetric loop-like topology. Our simulations confirm that this setup corresponds to the global minimum for particle escape, where the closed magnetic field lines effectively suppress outflows in favor of accretion and trapping. 
\end{itemize}

These results demonstrate that the inclination of the external field can strongly impact the escape channels of ionized particles from the disk. The analysis also shows high sensitivity of particle dynamics to both the strength and orientation of the magnetic field. 

\section{Astrophysical relevance} \label{sec:astro} 

The superposition of an intrinsic magnetosphere with an external, potentially misaligned magnetic field is not merely a theoretical curiosity but a likely configuration in various astrophysical environments. We identify two example scenarios where such interactions are particularly relevant: stellar-mass black hole binaries, where the external field is provided by a magnetized companion (such as a neutron star), and SMBHs, such as Sgr~A*, interacting with galactic magnetic fields or magnetized winds from nearby stars.

\subsection{Stellar-mass compact binaries} 

We first consider the model in stellar-mass black hole binaries, focusing particularly on microquasars. 
In these systems, the black hole accretes matter from a companion star—either via Roche-lobe overflow or by capturing material from a dense stellar wind \cite{2002apa..book.....F}. While the intrinsic field ($P$) is generated by the currents within the accretion disk or the plunging region, a significant external field ($B_{\rm ext}$) can be imposed by the magnetosphere of a neutron star companion or the magnetized wind of a massive companion. 

In a binary system, the orientation of the external field vector relative to the black hole spin varies with the orbital phase. As shown in Eq. (\ref{eq:Flux-general}), the magnetic flux is modulated by the azimuthal angle $\beta$. Consequently, as the companion star orbits the black hole, the effective inclination of the ambient field changes, leading to a periodic modulation of the jet power. On the other hand, the inner accretion flow with intrinsic field $P$ may undergo Lense-Thirring precession if misaligned with the orbital plane \cite{2025arXiv250320577B}. This precession modulates the polar inclination angle $\alpha$ on timescales much longer than the orbital period. Consequently, the periodic variation of the net magnetic flux $\Phi(\alpha, \beta)$ offers a purely geometric interpretation for the super-orbital modulation of the jet power. We suggest that this interplay naturally imprints variability at both orbital and precession timescales, contributing to the phenomenology of low-frequency QPOs observed in microquasars \cite{2011A&A...531A..59T,2021ApJ...906...92S,2022PASJ...74.1220S}.

Furthermore, microquasars frequently cycle between radio-loud (jetted) and radio-quiet states \cite{2005MNRAS.361..633N}. If the external field orientation is anti-aligned with respect to the intrinsic field such that the condition $P \approx -\frac{1}{2} r_H^2 B_z$ is met, the net magnetic flux threading the horizon vanishes. In this case, the cancellation of the magnetic flux threading the horizon suppresses jet formation even at high accretion rates, providing a geometric reasoning for the transition to the radio-quiet state. 

This implies that the transition to a radio-quiet state requires the external magnetic pressure (supplied by the companion's wind or magnetosphere) to be comparable to the pressure of the intrinsic field. Assuming the intrinsic field corresponds to the equipartition pressure in the inner disk ($B_{\rm int} \sim 10^7$--$10^8$ G for a typical Galactic $10 M_\odot$ black hole \cite{Daly:APJ:2019:}), this condition favors tight binary separations with a neutron star companion, or significant wind compression to achieve the necessary external field strength.

The superposition of fields creates the radial null rings at $\theta = 60^{\circ}$ (derived after Eq. (\ref{Eq:flux-alligned})). As inhomogeneities in the wind or disk cross these static null surfaces, they provide plausible sites for magnetic reconnection. We therefore suggest that the characteristic timescale of any resulting X-ray flares is set by the light-crossing time of the null region, $\tau \approx r_{\rm null}/c \sim r_H/c$, implying millisecond variability for a $10M_{\odot}$ black hole. 

Finally, the topological rearrangement of field lines during transitions between the hard state (typically jet-producing and radio-bright) and the soft state (typically jet-quenched and radio-faint) should be verifiable via polarimetry (e.g., IXPE). As the net flux approaches zero, the dominant field component shifts from poloidal to toroidal/horizontal, 
potentially producing a rotation of the polarization angle and a reduction in polarization degree near epochs of jet quenching, depending on the dominant X-ray emission site and field ordering.

\subsection{SMBHs: missing jet of Sgr~A* }

A suggested superposition model is also applicable to SMBHs, where the external magnetic field is supplied by the large-scale magnetic environment of the galactic nucleus. A prime candidate for this scenario is the SMBH at the center of our Galaxy, Sgr A*.

Unlike stellar-mass binaries, Sgr A* is embedded in the nuclear star cluster and interacting with the circum-nuclear disk. Radio observations reveal complex magnetic field structure in central region of the the Galaxy: the field within the dense clouds of the Central Molecular Zone (CMZ) is predominantly parallel to the Galactic plane \cite{2015llg..book..391M}. According to the Event Horizon Telescope (EHT) results \cite{2022ApJ...930L..16E}, the spin axis of Sgr~A* appears to be tilted toward us at a very small angle, lying almost in the plane of the Galaxy. While the internal magnetic field of Sgr A* is expected to align with the spin axis of the black hole, the Galactic magnetic fields are known to follow the spiral arms of the Galaxy, spiraling inward toward the Galactic center \cite{2013pss5.book..641B}. Under such conditions, an anti-aligned configuration between the Galactic magnetic field and the internal field of Sgr A* is not only possible but likely.  
Applying our results to Sgr A* provides a geometric framework for addressing the long-standing 'missing jet' problem and the system's anomalously low luminosity.

Despite its mass ($M \approx 4 \times 10^6 M_{\odot}$), Sgr A* is under-luminous and lacks a prominent, stable relativistic jet comparable to that of many AGNs. While the lack of a large-scale jet is often attributed to radiatively inefficient accretion or poor outflow collimation, it has also been suggested that a jet could exist but remain undetected due to instrumental limitations and interstellar scattering \cite{2024ApJ...974..116C}. Beyond these standard interpretations, we propose that the interaction with the ambient Galactic magnetic field offers a geometric explanation: specifically, that the misalignment of field components may physically stifle the formation of a prominent jet.  

If the orientation of the large-scale Galactic magnetic field is anti-aligned with the black hole's intrinsic magnetosphere, the system may be in a permanent or semi-permanent state of flux cancellation. In this scenario, the external vertical field effectively chokes the formation of a globally organized BZ jet, confining the activity to weak outflows or localized flares. 

Quantitatively, our model constrains the characteristic radius for jet dissipation to the range of $140 - 4500 \, r_g$ (where $r_g = GM/c^2$). At the distance of Sgr A*, this corresponds to an angular scale of $0.7 - 2.2$ mas. This radius marks the zone where the external Galactic field begins to dominate the intrinsic magnetosphere; the resulting formation of magnetic null points and field reversals disrupts the collimation of the outflow, effectively halting the jet's propagation before it can reach large scales. The estimate is derived from the condition of magnetic pressure balance between the intrinsic and external fields, adopting intrinsic split-monopole magnetic field with the horizon-scale strength of 10 G for Sgr~A*, and Galactic ambient field ranging from $\sim \mu\text{G}$ to $\sim \text{mG}$.


\section{Summary and Conclusions} \label{sec:dis_con} 

In this work, we presented an analytical model for the magnetosphere of a Schwarzschild black hole immersed in an oblique external magnetic field. 
By superposing the axisymmetric Blandford--Znajek split-monopole solution with the Bičák--Janiš solution for an inclined uniform field, we have investigated how magnetospheric topology, horizon-threading flux, and particle dynamics are modified by the misalignment of ambient magnetic sources.

Our main results are summarized as follows: 
\begin{itemize}
\setlength{\itemsep}{2pt}
  \setlength{\parskip}{2pt}
  \setlength{\parsep}{2pt}
\item We demonstrated that the superposition of intrinsic and external fields generically leads to the formation of magnetic null points and complex, hybrid magnetic topologies. Unlike the purely axisymmetric aligned cases, the introduction of a non-zero inclination angle breaks the symmetry of the magnetosphere, creating regions where field lines form closed loops or transition between paraboloidal and toroidal geometries. We provided analytical expressions for the location of these null points, which serve as potential sites for magnetic reconnection and particle acceleration. 
\item 
We derived an exact formula for the magnetic flux threading an arbitrary hemisphere of the event horizon. We showed that the flux is not constant but is strongly modulated by the azimuthal orientation ($\cos\beta$) of the external field. For anti-aligned vertical components ($B_z P<0$), we found magnetic null points above/below the equatorial region and derived closed-form expressions for their location $(r_{\rm null},\theta_{\rm null},\phi_{\rm null})$. Crucially, we identified a condition ($P \approx -0.5 r_H^2 B_z$), under which the net magnetic flux threading the horizon vanishes, providing a purely geometric mechanism for suppressing relativistic jets, even in systems with high accretion rates. 
\item 
We studied particle acceleration capability in this configuration by simulating a large number of trajectories starting from random positions in the Keplerian accretion disk. Our results show that the ability of particles to escape depends critically on the orientation of the external magnetic field. In the strictly anti-aligned case, the magnetic field forms a closed "trap" that prevents outflows and forces particles to accrete. However, tilting the external field breaks the symmetry and creates chaotic paths, which allow particles to escape. Consequently, we find that the number of escaping particles is actually highest when the field is slightly inclined, rather than perfectly aligned. 

\item 
We applied our geometric model to relevant astrophysical systems. For stellar-mass binaries, we proposed that the orbital modulation of the external field inclination leads to periodic variations in jet power, potentially contributing to low-frequency QPOs and  transitions between radio-loud to radio-quiet states. For Sgr A*, we argued that an anti-aligned orientation with respect to the Galactic magnetic field provides a natural geometric explanation for the ``missing jet'' problem, estimating a jet dissipation radius of $\sim 140$--$4500\,GM_{\rm SgrA^*}/c^2$,  where the outflow is quenched by the ambient Galactic magnetic field.
\end{itemize} 

Our results are in excellent agreement with the recent 3D GRPIC simulations by Figueiredo et al. \cite{refId0}, which investigated a similar inclined magnetic field configuration. Their fully kinetic study supports two of our primary findings: first, that the jet-supporting magnetic structure is significantly weakened as the external field inclination increases; and second, that despite this weakening, particle acceleration remains highly efficient even for high inclinations.

While 3D GRPIC simulations provide the most complete kinetic description of black hole magnetospheres, they are computationally expensive, limiting their ability to systematically test large parameter spaces. Our semi-analytic approach complements these studies by isolating the geometric effects of magnetic field superposition, enabling controlled and inexpensive exploration of magnetic inclination, field strength, and topology, and offering direct insight into the conditions under which jet-supporting configurations are weakened while particle acceleration persists. 
In the numerical tests performed in this work, collective plasma effects arising from particle–particle interactions and their back-reaction on the background magnetic field were neglected. Consequently, our many-particle simulations can be interpreted within a test-particle framework that isolates the geometric influence of the combined magnetic field on particle dynamics.

Our findings suggest that the interaction between intrinsic and ambient magnetic fields, largely overlooked in standard analyses, can be an important factor in determining the phenomenology of black hole systems. More broadly, the purely geometric nature of the model suggests that similar magnetospheric interactions should arise in a wide range of black hole systems embedded in structured or evolving ambient magnetic environments \cite{2024evn..conf....1T}. It would be interesting to extend this analysis to the Kerr metric in future work in order to fully incorporate frame-dragging effects and the electric fields induced by black hole rotation, which are essential for a self-consistent treatment of energy extraction and particle acceleration.

\begin{acknowledgements}
This work was partially supported by the Ministry of Education and Science (MES) of the Republic of Kazakhstan (RK), Grant No.{AP23489541}, and the Czech Science Foundation Grant (GA\v{C}R) No.~\mbox{23-07043S}. 
\end{acknowledgements}



\begin{thebibliography}{10}

\bibitem{2021ApJ...910L..13E}
{Event Horizon Telescope Collaboration}.
\newblock {First M87 Event Horizon Telescope Results. VIII. Magnetic Field Structure near The Event Horizon}.
\newblock {\em \apjl}, 910(1):L13, March 2021.

\bibitem{2024ApJ...964L..26E}
{Event Horizon Telescope Collaboration}.
\newblock {First Sagittarius A* Event Horizon Telescope Results. VIII. Physical Interpretation of the Polarized Ring}.
\newblock {\em \apjl}, 964(2):L26, April 2024.

\bibitem{Bla-Zna:1977:MNRAS:}
R.~D. {Blandford} and R.~L. {Znajek}.
\newblock {Electromagnetic extraction of energy from Kerr black holes}.
\newblock {\em \mnras}, 179:433--456, May 1977.

\bibitem{2024PhRvD.109f3005K}
Saltanat {Kenzhebayeva}, Saken {Toktarbay}, Arman {Tursunov}, and Martin {Kolo{\v{s}}}.
\newblock {Black hole in a combined magnetic field: Ionized accretion disks in the jetlike and looplike configurations}.
\newblock {\em \prd}, 109(6):063005, March 2024.

\bibitem{Kom:2004:MNRAS:}
S.~S. {Komissarov}.
\newblock {Electrodynamics of black hole magnetospheres}.
\newblock {\em \mnras}, 350(2):427--448, May 2004.

\bibitem{Tch-Nar-McK:2010:APJ:}
Alexander {Tchekhovskoy}, Ramesh {Narayan}, and Jonathan~C. {McKinney}.
\newblock {Black Hole Spin and The Radio Loud/Quiet Dichotomy of Active Galactic Nuclei}.
\newblock {\em \apj}, 711(1):50--63, Mar 2010.

\bibitem{refId0}
{Figueiredo, Enzo}, {Cerutti, Benoît}, and {Parfrey, Kyle}.
\newblock Effect of magnetic field inclination on black hole jet power and particle acceleration.
\newblock {\em A\&A}, 700:L19, 2025.

\bibitem{1985MNRAS.212..899B}
J.~{Bicak} and V.~{Janis}.
\newblock {Magnetic fluxes across black holes}.
\newblock {\em \mnras}, 212:899--915, February 1985.

\bibitem{Kom-McK:2007:MNRAS:}
S.~S. {Komissarov} and Jonathan~C. {McKinney}.
\newblock {The `Meissner effect' and the Blandford-Znajek mechanism in conductive black hole magnetospheres}.
\newblock {\em \mnras}, 377(1):L49--L53, May 2007.

\bibitem{Kol-Jan:2020:RAG:}
M.~{Kolo{\v s}} and A.~Janiuk.
\newblock {Simulations of black hole accretion torus in various magnetic field configurations}.
\newblock In S.~{Hled{\'{\i}}k} and Z.~{Stuchl{\'{\i}}k}, editors, {\em RAGtime 20-22: Workshops on black holes and neutron stars}, page 153–164, December 2020.

\bibitem{Par-Phi-Cer:2019:PRL:}
Kyle {Parfrey}, Alexander {Philippov}, and Beno{\^\i}t {Cerutti}.
\newblock {First-Principles Plasma Simulations of Black-Hole Jet Launching}.
\newblock {\em \prl}, 122(3):035101, January 2019.

\bibitem{Nak-etal:2018:APJ:}
M.~{Nakamura}, K.~{Asada}, K.~{Hada}, H.~{Pu}, S.~{Noble}, C.~{Tseng}, K.~{Toma}, M.i {Kino}, H.~{Nagai}, K.~{Takahashi}, J.-C. {Algaba}, M.~{Orienti}, K.~{Akiyama}, A.~{Doi}, G.~{Giovannini}, M.~{Giroletti}, M.~{Honma}, S.~{Koyama}, R.~{Lico}, K.~{Niinuma}, and F.~{Tazaki}.
\newblock {Parabolic Jets from the Spinning Black Hole in M87}.
\newblock {\em \apj}, 868(2):146, Dec 2018.

\bibitem{Kol-Sha-Tur:2023:EPJC:}
Martin {Kolo{\v{s}}}, Misbah {Shahzadi}, and Arman {Tursunov}.
\newblock {Charged particle dynamics in parabolic magnetosphere around Schwarzschild black hole}.
\newblock {\em European Physical Journal C}, 83(4):323, April 2023.

\bibitem{2000ApJ...533L.115L}
Li-Xin {Li}.
\newblock {Extracting Energy from a Black Hole through Its Disk}.
\newblock {\em \apjl}, 533(2):L115--L118, April 2000.

\bibitem{2007MNRAS.374..647W}
Ding-Xiong {Wang}, Yong-Chun {Ye}, Yang {Li}, and Dong-Mei {Liu}.
\newblock {A toy model for magnetic connection in black hole accretion disc}.
\newblock {\em \mnras}, 374(2):647--656, January 2007.

\bibitem{Tur-Dad:2019:Universe:}
Arman {Tursunov} and Naresh {Dadhich}.
\newblock {Fifty Years of Energy Extraction from Rotating Black Hole: Revisiting Magnetic Penrose Process}.
\newblock {\em Universe}, 5(5):125, May 2019.

\bibitem{2008ApJ...681...96H}
Tao {Hu} and Qiuhe {Peng}.
\newblock {Nucleosynthesis in Accretion and Outflow Regions around Black Holes}.
\newblock {\em \apj}, 681(1):96--103, July 2008.

\bibitem{2017Galax...5...15J}
Agnieszka {Janiuk}, Michal {Bejger}, Petra {Sukova}, and Szymon {Charzynski}.
\newblock {Black Hole Accretion in Gamma Ray Bursts}.
\newblock {\em Galaxies}, 5(1):15, February 2017.

\bibitem{Tur-etal:2019:ApJ:}
Arman {Tursunov}, Zden{\v{e}}k {Stuchl{\'\i}k}, Martin {Kolo{\v{s}}}, Naresh {Dadhich}, and Bobomurat {Ahmedov}.
\newblock {Supermassive Black Holes as Possible Sources of Ultrahigh-energy Cosmic Rays}.
\newblock {\em \apj}, 895(1):14, May 2020.

\bibitem{Stu-Vrb:2025:Ent:}
Zden{\v{e}}k {Stuchl{\'\i}k} and Jaroslav {Vrba}.
\newblock {Ionized Keplerian Disks Demonstrating Interplay Between Strong Gravity and Magnetism}.
\newblock {\em Entropy}, 27(12):1253, December 2025.

\bibitem{2002apa..book.....F}
Juhan {Frank}, Andrew {King}, and Derek~J. {Raine}.
\newblock {\em {Accretion Power in Astrophysics: Third Edition}}.
\newblock 2002.

\bibitem{2025arXiv250320577B}
D.~A. {Bollimpalli}, J.~{Hor{\'a}k}, W.~{Klu{\'z}niak}, and P.~C. {Fragile}.
\newblock {Misalignment of the Lense-Thirring precession by an accretion torque}.
\newblock {\em arXiv e-prints}, page arXiv:2503.20577, March 2025.

\bibitem{2011A&A...531A..59T}
G.~{T{\"o}r{\"o}k}, A.~{Kotrlov{\'a}}, E.~{{\v{S}}r{\'a}mkov{\'a}}, and Z.~{Stuchl{\'\i}k}.
\newblock {Confronting the models of 3:2 quasiperiodic oscillations with the rapid spin of the microquasar GRS 1915+105}.
\newblock {\em \aap}, 531:A59, July 2011.

\bibitem{2021ApJ...906...92S}
Krista~Lynne {Smith}, Celia~R. {Tandon}, and Robert~V. {Wagoner}.
\newblock {Confrontation of Observation and Theory: High-frequency QPOs in X-Ray Binaries, Tidal Disruption Events, and Active Galactic Nuclei}.
\newblock {\em \apj}, 906(2):92, January 2021.

\bibitem{2022PASJ...74.1220S}
Zden{\v{e}}k {Stuchl{\'\i}k}, Martin {Kolo{\v{s}}}, and Arman {Tursunov}.
\newblock {Large-scale magnetic fields enabling fitting of the high-frequency QPOs observed around supermassive black holes}.
\newblock {\em \pasj}, 74(5):1220--1233, October 2022.

\bibitem{2005MNRAS.361..633N}
Carlo {Nipoti}, Katherine~M. {Blundell}, and James {Binney}.
\newblock {Radio-loud flares from microquasars and radio-loudness of quasars}.
\newblock {\em \mnras}, 361(2):633--637, August 2005.

\bibitem{Daly:APJ:2019:}
Ruth~A. {Daly}.
\newblock {Black Hole Spin and Accretion Disk Magnetic Field Strength Estimates for More Than 750 Active Galactic Nuclei and Multiple Galactic Black Holes}.
\newblock {\em \apj}, 886(1):37, November 2019.

\bibitem{2015llg..book..391M}
Mark~R. {Morris}.
\newblock {Manifestations of the Galactic Center Magnetic Field}.
\newblock In Kenneth {Freeman}, Bruce {Elmegreen}, David {Block}, and Matthew {Woolway}, editors, {\em Lessons from the Local Group: A Conference in honor of David Block and Bruce Elmegreen}, page 391. 2015.

\bibitem{2022ApJ...930L..16E}
{Event Horizon Telescope Collaboration}.
\newblock {First Sagittarius A* Event Horizon Telescope Results. V. Testing Astrophysical Models of the Galactic Center Black Hole}.
\newblock {\em \apjl}, 930(2):L16, May 2022.

\bibitem{2013pss5.book..641B}
Rainer {Beck} and Richard {Wielebinski}.
\newblock {Magnetic Fields in Galaxies}.
\newblock In Terry~D. {Oswalt} and Gerard {Gilmore}, editors, {\em Planets, Stars and Stellar Systems. Volume 5: Galactic Structure and Stellar Populations}, volume~5, page 641. 2013.

\bibitem{2024ApJ...974..116C}
Erandi {Chavez}, Sara {Issaoun}, Michael~D. {Johnson}, Paul {Tiede}, Christian {Fromm}, and Yosuke {Mizuno}.
\newblock {Prospects of Detecting a Jet in Sagittarius A* with Very-long-baseline Interferometry}.
\newblock {\em \apj}, 974(1):116, October 2024.

\bibitem{2024evn..conf....1T}
Arman {Tursunov} and Silke {Britzen}.
\newblock {Ambient magnetic field interactions influencing jet propagation}.
\newblock In E.~{Ros}, P.~{Benke}, S.~A. {Dzib}, I.~{Rottmann}, and J.~A. {Zensus}, editors, {\em Proceedings of the 16th EVN Symposium}, pages 1--6, September 2024.

\end{thebibliography}

\providecommand{\noopsort}[1]{}\providecommand{\singleletter}[1]{#1}

\end{document}